**Bonding Properties of Manganese Nitrides at High Pressure and the Discovery of $MnN_4$ with planar $N_4$ rings**

Li Li[1], Kuo Bao[1]* Xingbin Zhao[1], and Tian Cui[2,1]*

[1]State Key Laboratory of Superhard Materials, College of Physics, Jilin University, Changchun, 130012, China

[2]Institute of High Pressure Physics, School of Physical Science and Technology, Ningbo University, Ningbo 315211, China;

*Corresponding authors: baokuo@jlu.edu.cn, and cuitian@nbu.edu.cn

## Abstract

It is not easy to synthesize high quality manganese nitrides experimentally; however, owing to the intrinsic characteristics of manganese and nitrogen, these materials possess remarkable properties and have highly promising applications. In this study, we systematically examined the stoichiometric phase spaces of Mn−N compounds from 0 to 100 GPa using *ab initio* calculations and constructed a high-pressure magnetic phase diagram. Remarkably, N-rich $MnN_4$ with a planar $N_4$ ring was discovered for the first time in the pressure range of 40–100 GPa. The electronic structures revealed that the $N_4$ ring is formed by the $sp^2$ hybridization of nitrogen atoms. Furthermore, its superconducting transition temperature is approximately 1.6 K, and its bulk modulus is 381 GPa, thereby rendering it a potential hard superconductive material. Moreover, we determined a new phase transition sequence for MnN: the semi-conducting non-magnetic (NM) *zb* phase (5 GPa) first transforms to metallic antiferromagnetic (AFM) *NiAs* (40 GPa), which further transforms to the more stable metallic ferromagnetic (FM) *rs* phase. The mechanical properties indicated that covalent interactions have a significant effect on the hardness of the N-rich structures and almost no effect on the Mn-rich structures in Mn−N compounds. Our work provides an overview of the Mn−N compounds and their properties under pressure and presents an updated phase diagram.

# Introduction

Manganese is a unique transition metal (TM), not only because it has the highest magnetic moment among all the elements, but also because it has one of the highest electron densities. Thus, its compounds have also drawn considerable attention owing to their magnetic, electronic, semi conductive, superconductive, and photovoltaic behaviors. Parallelly, nitrogen shows interesting structures and behavior under high pressure (HP).[1, 2] Thus, manganese nitrides possess remarkable properties and have promising applications. However, the synthesis of "pure" TM nitrides or compounds with an integral ratio is a challenge using the conventional methods, because nitrogen (N ≡ N) is highly stable and unreactive. The use of HP is an effective and clean method to change hybridization modes, bonding ways, and valence electron orbitals and to synthesize novel materials with special physicochemical properties. A series of poly-nitrogen forms have been predicted and synthesized at HP.[3-10] In addition to high energy density materials, TM nitrides have been drawing considerable attention owing to their outstanding mechanical,[11] superconductive, and magnetic properties under certain conditions. TiN was synthesized as a superconducting film that underwent disorder-driven superconductor–insulator transition.[12] NbN[13, 14] and MoN,[15] both of which are ultra-hard and superconducting unconventional functional nitrides, have been reported. Several studies have shown that different bonding patterns between nitrogen atoms have significant influence on the mechanical properties of materials. In the Fe−N system, the hardness of $Fe_3N_8$ and $FeN_3$ with N chains was measured to be above 30 GPa.[16] Recently, trigonal $WN_6$ with armchair-like hexazine ($N_6$) rings was synthesized, and its hardness was determined to be 57.9 GPa by theoretical calculations.[17] The $N_2$ units in $CrN_2$, $TiN_2$, $IrN_2$ and $OsN_2$ have been shown to impart ultra-incompressible properties with high bulk modulus (B) and shear modulus (G).[11, 18]

Mn-N compounds manifest rich magnetic phases. Until now, the ε-$Mn_4N$, ζ-$Mn_2N$, θ-MnN, and η-$Mn_3N_2$ phases have been synthesized [19-21]. Face-centered cubic ε-$Mn_4N$ exhibits interesting ferrimagnetic (FiM) order magnetism; the magnetic moment of Mn on the corner (Mn-I) is 3.3 $\mu_B$ and antiparallel to the three Mn (Mn-II) on face center is 0.7 $\mu_B$ [22]. It exhibits a distinct anomalous Hall Effect in perpendicular magnetic fields [23, 24] and shows an extremely high Curie temperature of ~710 K.[25] ζ-$Mn_2N$ has been shown to form ζ-$Fe_2N$ type AFM orthorhombic phase, and a neutron diffraction experiment confirmed the space group (*SG*) to be *Pbcn* with lattice constants a= 5.668 Å, b = 4.909 Å, and c= 4.537 Å.[26] Later, the $P6_3/mmc$-$Mn_2N_{1.06}$ was prepared with an In-Na flux at 700 °C and 5 MPa of $N_2$.[20] η-$Mn_3N_2$, a tetragonal (*I*4/*mmm* ) metal with lattice constants a = b = 2.994 Å, and c = 12.499 Å [27] and a Nèel temperature of 925K has also been demonstrated.[28] Its magnetic moment is parallelly aligned in the (001) plane, lies along the [100] direction, and presents an AFM order along [001]. Due to its high reversible capacity and good cycle performance, it could be used as an anode material for lithium-ion batteries.[29] However, the ground state structure of MnN has long been discussed. θ-MnN exists experimentally in a tetragonally distorted rock salt (*rs*) structure with AFM order. It exhibits metallic properties with lattice parameters c/a = 0.984 in the [001] direction, and the magnetic moment of Mn atom is 3.3 $\mu_B$ at 0 K.[30] However, many theories were proposed for the hypothetical cubic zinc-blende (*zb*). A. Janotti *et al* [31] explained why the ground state *zb*-AFM state has a larger lattice constant and higher magnetic moment of the Mn atom than the FM phase. Hong *et al.* studied MnX (X= N, P, As, Sb) binary compounds in the *NiAs* and *zb* phases, and reported that their ground state is the AFM *zb* phase, because the energy of the N-*p* level is much lower than other anions *p*, and it is more likely to form a lower coordination number *zb* phase. Meanwhile, their AFM orders are larger, which reduces *p-d* and *d-d* coupling; consequently, the AFM state is more stable.[32] However, compared with iron nitrides, which easily form N-rich Fe−N compounds under HP, such as $FeN_2$, $FeN_4$, $Fe_3N_8$, the N-rich Mn−N compounds are also worth investigating.

In this study, we investigated Mn−N compounds from 0 to 100 GPa based on *ab initio* calculations. A series of new N-rich magnetic phases were predicted. Remarkably, N-rich $MnN_4$ with a planar $N_4$ ring was discovered, and its bonding properties were analyzed, indicating that it could constitute a potentially hard superconductive material. Moreover, the phase transition sequence for MnN was updated. Finally, the incompressibility of all the structures of the Mn−N compounds was determined.

## COMPUTATIONAL METHODS AND DETAILS

The structure search for Mn–N compounds were conducted based on evolutionary algorithm using the variable-composition mode of the USPEX code[33, 34] at 0, 50, and 100 GPa. The first generation was produced randomly, and the succeeding generations were obtained by applying 50% heredity, 20% symmetric, 10% lattice mutation, and 20% soft mutation. Afterward, we fixed the ratios and varied the pressures to conduct the structural search again. Structure cells with Mn:N ratios of 3:1, 2:1, 3:2, 1:1, 1:2, and 1:4 within 1−4 formula units (f.u.) were obtained. Structural relaxations, electronic properties, and total energies were calculated using density functional theory (DFT) calculations implemented with the Vienna Ab initio Simulation Package (VSAP)[35] with the project augmented wave method.[36] Additionally, the Perdew-Burke-Ernzerhof parametrization of the generalized gradient approximation was used.[37] A cutoff energy of 850 eV was chosen. Mokhorst-Pack k-grid meshes[38] with a reciprocal space resolution of $2\pi \times 0.03$ Å$^{-1}$ and $2\pi \times 0.02$ Å$^{-1}$ were used for determining the thermodynamic parameters and electronic properties, respectively. The electronic configurations of Mn and N were chosen as $3d^6 4s^1$ and $2s^2 2p^3$, respectively. Phonon dispersion curves were obtained using the PHONOPY code[39] based on a supercell approach with force-constant matrices and electron-phonon coupling (EPC) were performed by density functional perturbation theory as implemented in the Quantum ESPRESSO package.[40] Norm-conserving potentials were used with a kinetic energy cutoff of 90 Ry. The q-point mesh of the electron-phonon interaction matrix element adopted $4 \times 4 \times 4$ in $MnN_4$. Elastic constants were calculated using the strain stress method, while bulk modulus (B) and shear modulus (G) were derived from the Voigt–Reuss–Hill averaging scheme.[41]

## Results and discussion

### 3.1. Phase Stability of Mn−N Compounds at High Pressure.

Formation enthalpies ($\Delta H_f$) from the global energy minimum of the Mn−N system were computed from the convex hull. $\Delta H_f$ of each structure was calculated using the following relation: $\Delta H_f(Mn_xN_y) = [H(Mn_x N_y) - xH(Mn) - yH(N)]/(x+y)$ ($x, y = 0, 1, 2,\ldots$). The α-Mn and low-energy N phases (α, *Pbcn*, *P*2/*c*, and cg phases) at different pressures were used as the reference phases.[42] From the **Figure 1**a, at 0 GPa, only $Mn_3N$, $Mn_2N$ and MnN are the stable phases. The experimentally determined the cubic $Mn_4N$ phase[23, 43] is always above the convex hull. The lattice constants are listed in **Table S1**. In terms of structures of Mn−N compounds, with increasing nitrogen content, the hexagonal layers composed of edge-sharing $NMn_6$ tri-prisms in Mn-rich structures transform into polyhedral structures centered around the Mn atom in N-rich structures (**Figure S1**). The metastable phases (dynamically and mechanically stable phases, namely, NM *C*2/*m*-$Mn_3N$ and NM-*Cmmm*-$MnN_4$) by structure search are worth studying because these are possible to be synthesized experimentally, like the AFM *Pbcn*-$Mn_2N$.[4, 26] The corresponding structure diagrams are shown in **Figure S2**, and the phonon spectra are shown in **Figure S3** and **S4**. Three stable Mn-rich manganese polynitrides were predicted. For $Mn_2N$, experimental studies proposed a *Pbcn* phase. Examination on the relative enthalpy (**Figure S5**) reveal that *P*6$_3$/*mmc*-$Mn_2N$, obtained by structure search, has a lower energy. *R*$\bar{3}$*m*-$Mn_3N_2$ is first predicted to be thermodynamically stable from 17 to 100 GPa, since *R*$\bar{3}$*m*-$Mn_3N_2$ is dynamically unstable below 10 GPa; the ground state is the experimentally determined AFM-*I*4/*mmm* phase at 0 GPa. *P*$\bar{6}$*m*2-$Mn_3N$ and *P*6$_3$/*mmc*-$Mn_2N$ are stable from 0 to 100 GPa. For $MnN_4$, the NM-*Immm* can be obtained by the decomposition of $MnN_2$ (**Figure 1**b), and it can remain stable at 40-100 GPa. From the phase diagram (**Figure 1**c), the NM-*zb* phase first transforms to AFM-*NiAs* at5 GPa, which further transforms to the more stable FM-*rs* phase at 40 GPa. Findings on $MnN_2$ will be reported in another work.

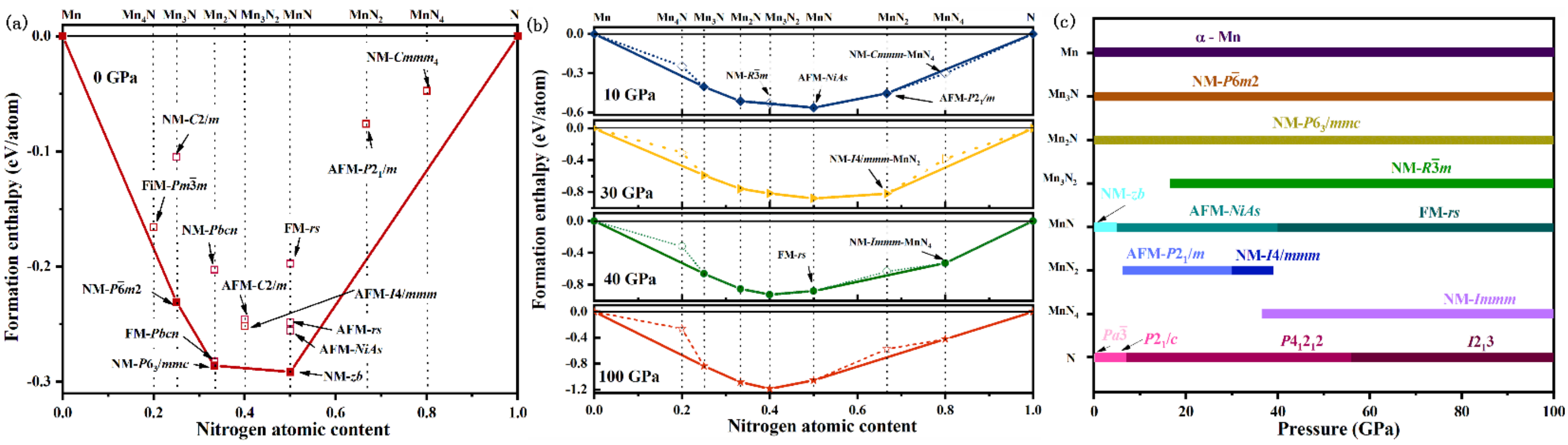

**Figure 1.** The magnetic phases (i.e., NM, FM, and AFM) of thermodynamic stability and the stable pressure ranges of Manganese nitrogen system. (a), (b) Convex hull diagram of the Mn–N system at different pressures at 0K. The solid squares connected by solid lines on the convex hull are represent thermodynamics and dynamics stable. (c) Pressure-composition phase diagram of the Mn–N system.

### 3.2. Bonding, Electronic, and Superconductive Properties of NM-*Immm*-$MnN_4$ Phase with $N_4$ Rings

Although NM-*Immm*-$MnN_4$ phase cannot be maintained at ambient pressure, the novel bonding mechanism of N atoms is worth studying. Its structure contains $MnN_6$ octahedrons that are connected by N atoms (**Figure 2**a). A series of poly-nitrogen forms were predicted and synthesized. Such as, the armchair-like hexazine ($N_6$) rings[17, 44] the pentazolate cyclo-$N_5^-$,[45-50] the $N_2$ and $N_6$ unit,[3, 4, 51] and other N-chains.[16, 52-55] Recently, 4-membered nitrogen ring was reported in *Cm*-$Al_2N_7$,[56] which is similar with the nitrogen rings in *Immm*-$MnN_4$. In $Al_2N_7$, there are two bond lengths of 1.36 Å and 1.37 Å respectively. And the bond lengths between N atoms are equal in $MnN_4$. In **Figure 2b**, the electron local function (ELF) of this phase is calculated for bonding analysis, and it is valid for determining the bonding characteristics, such as, covalent, metallic, and ionic bonds and lone pairs. The results of ELF show strong electronic localization between adjacent N1-N2 atoms within the planar $N_4$ rings with a scale larger than 0.75. This implies that covalent bonds in N1-N2 and the N atoms are in the *sp*$^2$ hybridization. The N1-N2 bond length in the $N_4$ rings is 1.380 Å, which is extremely close to the N-N single bond at 40 GPa. The Mn-N1 and Mn-N2 bond lengths are 1.930 Å and 1.807 Å, respectively.

Moreover, the crystal orbital Hamilton population COHP (ICOHP) was analyzed to characterize the bonding in $MnN_4$ as depicted in **Figure 2c**. The COHP plot of the N1−N2 pairs suggests that the bonding states are fully occupied in the energy range from −12 eV to −4 eV, while the antibonding states are partially occupied, thus forming covalent bonds between N1 and N2 of the $N_4$ ring. There are also some covalent interactions between Mn and N due to the partial occupation of the bonding states. The value of ICOHP reflects the bond strength; the interaction between N1−N2 is the strongest (−ICOHP = 12.024); that between Mn and N1 (−ICOHP = 4.101) and Mn and N2 (−ICOHP = 2.848) are next; and the interaction between Mn-Mn (−ICOHP = 0.681) is the weakest. Furthermore, the Laplacian of the charge density at the critical point was calculated to analyze the bonding behavior derived from the quantum theory of atoms in molecule model (QTAIM) (**Figure 2d**)[57]. The bond critical points (BCPs) between N1 and the nearest N2 possesses an electron density (ρ(r)) of 2.418 a.u. and a Laplacian ($\nabla^2\rho(r)$) of -20.255 a.u., which indicates stronger covalent interactions due to positive ρ(r) and negative $\nabla^2\rho(r)$. This result is in line with COHP and ELF. The BCP between Mn and the nearest N1 (N2) atom possesses a ρ(r) value of 0.678 a.u. (0.678 a.u.) and a $\nabla^2\rho(r)$ value of 14.321 a.u. (11.323 a.u.). The positive ρ(r) and $\nabla^2\rho(r)$ values imply that there is not entirely closed-shell interaction between Mn and N atoms, and that there are partial charge transfer bonds.[58] In fact, the Bader charge analysis revealed that each Mn atom loses ~1.2 $e^-$ to adjacent N atoms.

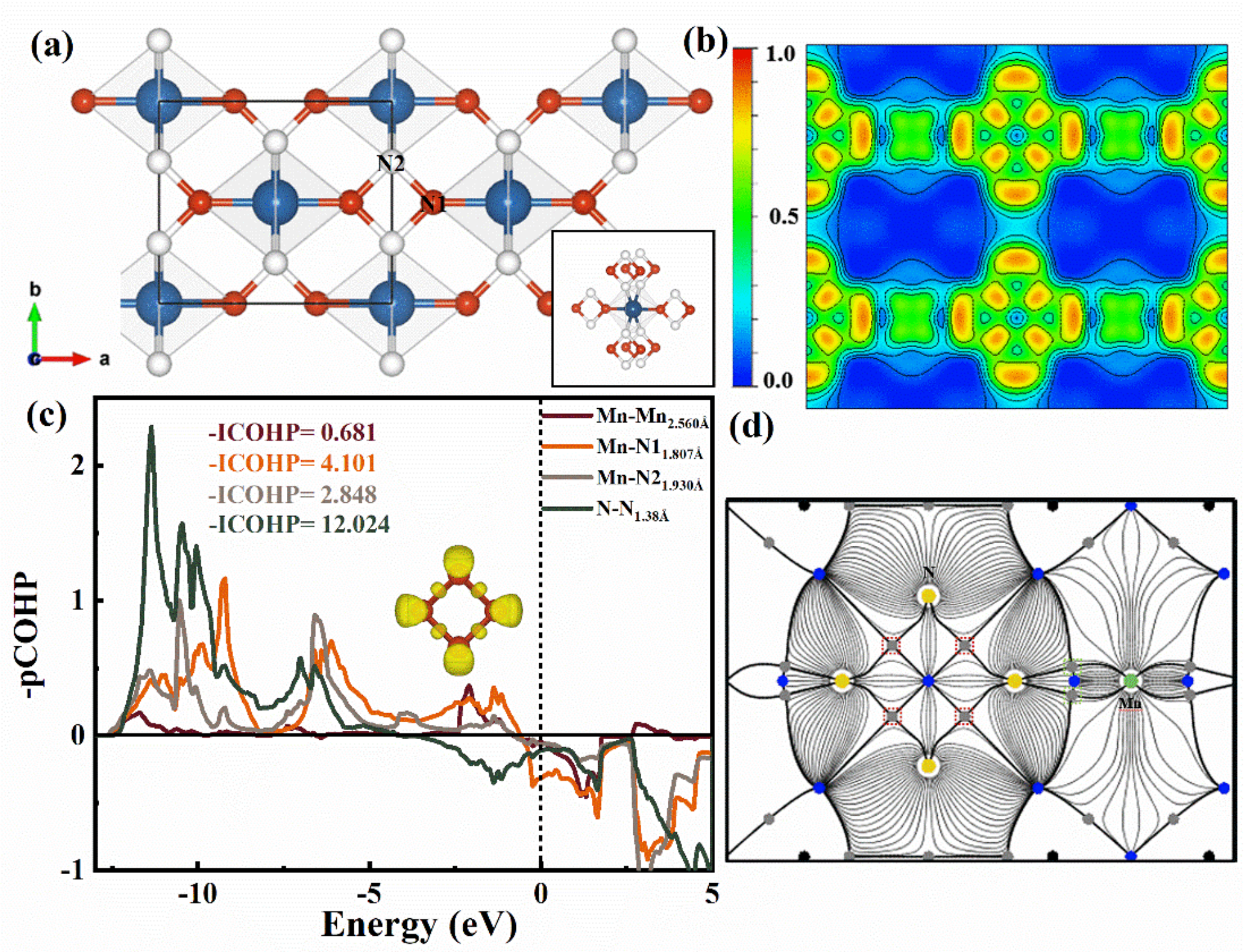

**Figure 2.** Crystal structures and electronic properties of $MnN_4$. (a) Crystal structure. Blue, red and white spheres denote Mn, N1 and N2 atoms. (b) The ELF. (c) The COHP, positive values represent bonding states, zero means nonbonding state, and negative values represent anti-bonding state. Inset represents the planer $N_4$ ring with an isosurface value of 0.8. (d) Gradient paths and critical points derived from QTAIM analysis for $MnN_4$, gray spheres represent BCPs, yellow spheres represent N, and green spheres represent Mn. Gray spheres in red dashed line boxes represent BCPs between N and N, in green dashed line boxes represent BCPs between Mn and N.

To understand the formation mechanism of the novel NM-*Immm*-$MnN_4$, the calculated DOS are shown in **Figure 3**a, the Mn-3*d* orbital plays a dominant role on the Fermi level and between the energy range from -13 eV to -2.5 eV, dominated by N-*p*. The s, $p_x$, and $p_y$ orbitals of N-*p* have the same peak energy, which indicates that these orbitals of each N atoms form three $sp^2$ hybridized orbitals for the $N_4$ rings, while the two $sp^2$ orbitals of each N atom form two σ bonds with the $sp^2$ hybridized orbitals of two neighboring N atoms. The extra $sp^2$ orbitals form lone pairs (see the ELF in inset of **Figure 2**c), while the four $p_z$ orbitals form delocalized π bonds (**Figure 3**b). The Mn-4*s* profile throughout this range of energy is very small, which indicates that charge transfer from Mn to N. Thus, the partially transferred electrons that filled the π-antibonding states impart metallicity and promote the mutual interaction between Mn and N, which results in a metallic state as well. This also confirmed the results of COHP.

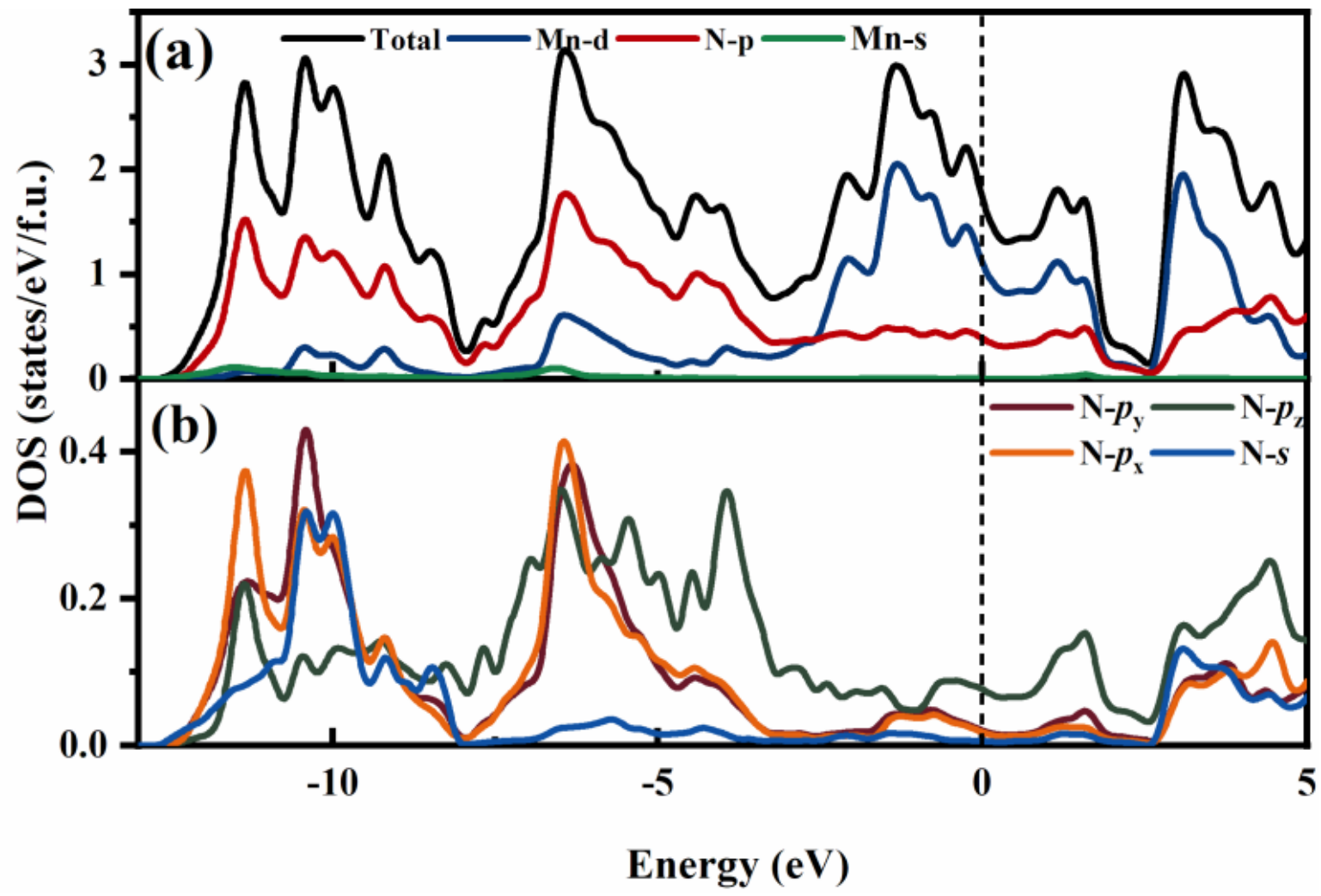

**Figure 3.** Projected DOS of NM-*Immm* at 40 GPa.

Due to the sizable DOS at the Fermi level of NM-*Immm*-$MnN_4$, we estimated the superconducting transition temperature ($T_c$) at 40 GPa. The phonon spectrum, PHDOS, and Eliashberg spectral function $\alpha^2F(\omega)$ together with the electron-phonon integral $\lambda$ of NM-*Immm* at 40 GPa are shown in **Figure 4**. The EPC parameters, namely, $\lambda \approx 0.41$ and $T_c \approx 1.6$ K were estimated using the McMillan equation modified by Allen and Dynes[59] with a Coulomb pseudopotential $\mu^* = 0.1$.

$$T_c = \frac{\omega_{\ln}}{1.2}\exp\left[-\frac{1.04(1+\lambda)}{\lambda-\mu^*(1+0.62\lambda)}\right] \quad (1)$$

NM-*Immm*-$MnN_4$ has weak-coupling superconductivity. Combined with the COHP and DOS, the interaction between Mn and N is the main contribution.

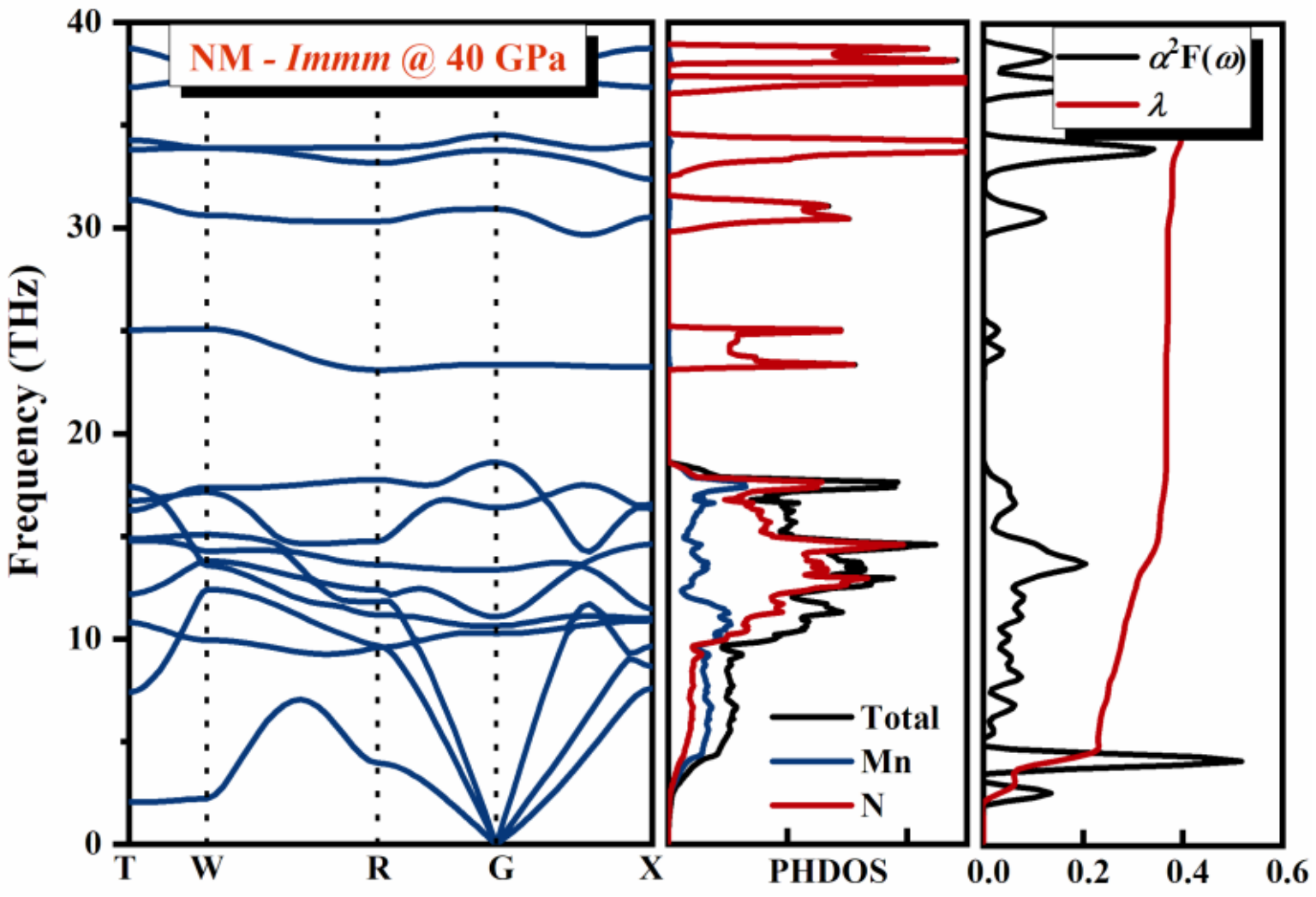

**Figure 4.** Phonon spectrum (left), PHDOS (middle), and Eliashberg spectral function $\alpha^2F(\omega)$ together with the electron-phonon integral $\lambda$ (right) of NM-*Immm* at 40 GPa.

### 3.3. Phase Transformation Sequence and Electronic Properties of MnN at High Pressure

Many theoretical studies on MnN have reported that the AFM *zb*-phase should be more energetically favorable than the *rs*-phase [32, 60, 61] similar to that observed in FeN and CoN. Miao *et al* suggested that *zb*-MnN could possibly be grown under N-rich conditions by controlled doping with S, Fe, and Co, and the transition from *zb* to the stable *rs* structure was observed with 4% compression.[61] However, there are no experimental reports on the *zb* phase in MnN until now. Additionally, the magnetic state of MnN continues to be discussed. Thus, we systematically studied MnN compounds from 0 to 100 GPa according to three structural stability criteria. The *zb*, *rs*, and *NiAs* structures with their magnetic phases were fully relaxed. For the experimental AFM-*rs* phase, lattice constants of a = 4.17 Å and c/a = 0.988 were observed in this study, which are within the experimental values reported in the literature.[30] This phenomenon of lattice collapse along the c axis is not observed in the FM or NM phases. The relative enthalpies are shown in **Figure 5**a. In terms of energy, the AFM-*zb* phase is the ground state. However, for phonon calculation (**Figure S4**), the AFM- and FM-*zb* phases are dynamically unstable, while there are no imaginary frequencies within the NM-*zb* phase. For the *rs* phase, both the FM- and AFM phases are stable at ambient pressure. Phonon spectrum reveals that the NM-*zb* phase is more stable under negative pressure (volume expansion) than under pressure. The phonon spectrum of the AFM-*zb* phase was also calculated under negative pressure; however, this phase was still unstable. To understand this phenomenon, the electronic properties were studied (**Figure 5**b, c and f). Remarkably, the results predict the existence of a small bandgap (0.1 eV) around the Fermi level of the NM-*zb* phase, which endows poor semiconductive properties. It is attributed to the splitting of the $e_g$. The $d_{x2-y2}$ orbital is repelled above the Fermi level, while the $d_{z^2}$ orbital is pushed below the Fermi level. However, the AFM-*zb*-phase exhibits electroconductibility due to magnetic interactions, which lead to the Mn-$e_g$ orbital contribute to the Fermi level. The N−N distance are 3.013 and 3.051 Å in NM and AFM *zb*-phase, respectively. They are much larger than single bond N−N. Thus, we only performed the projected crystal orbital Hamilton population analysis (COHP) of MnN (**Figure 5** d and e) to determine the structural stability. The integrated COHP (ICOHP) analysis reconfirms that the NM-*zb* phase (-ICOHP = 2.284) should be more stable than AFM- *zb* (-ICOHP = 2.041) owing to the higher bond strength. Under an increasing pressure of ~5 GPa, the NM-*zb* phase transforms into the AFM hexagonal *NiAs* metal. This is similar to that observed with FeN:[16, 62] the NM-*zb* phase of FeN transforms to the FM- *NiAs* phase at ~24 GPa. When the pressure is higher than 40 GPa, AFM-*NiAs* has an imaginary phonon frequency. Thus, the FM-*rs*-phase appears, exhibiting metallicity (**Figure S6**). Interestingly, the phonon topology can be observed in the phonon spectrum of AFM-*NiAs* (**Figure S3** h and i).

To summarize, the phase transition for MnN compounds proceeds in the following sequence:

semi-conducting NM-*zb* (5 GPa) → metallic AFM-*NiAs* (40 GPa) → metallic FM-*rs* (2)

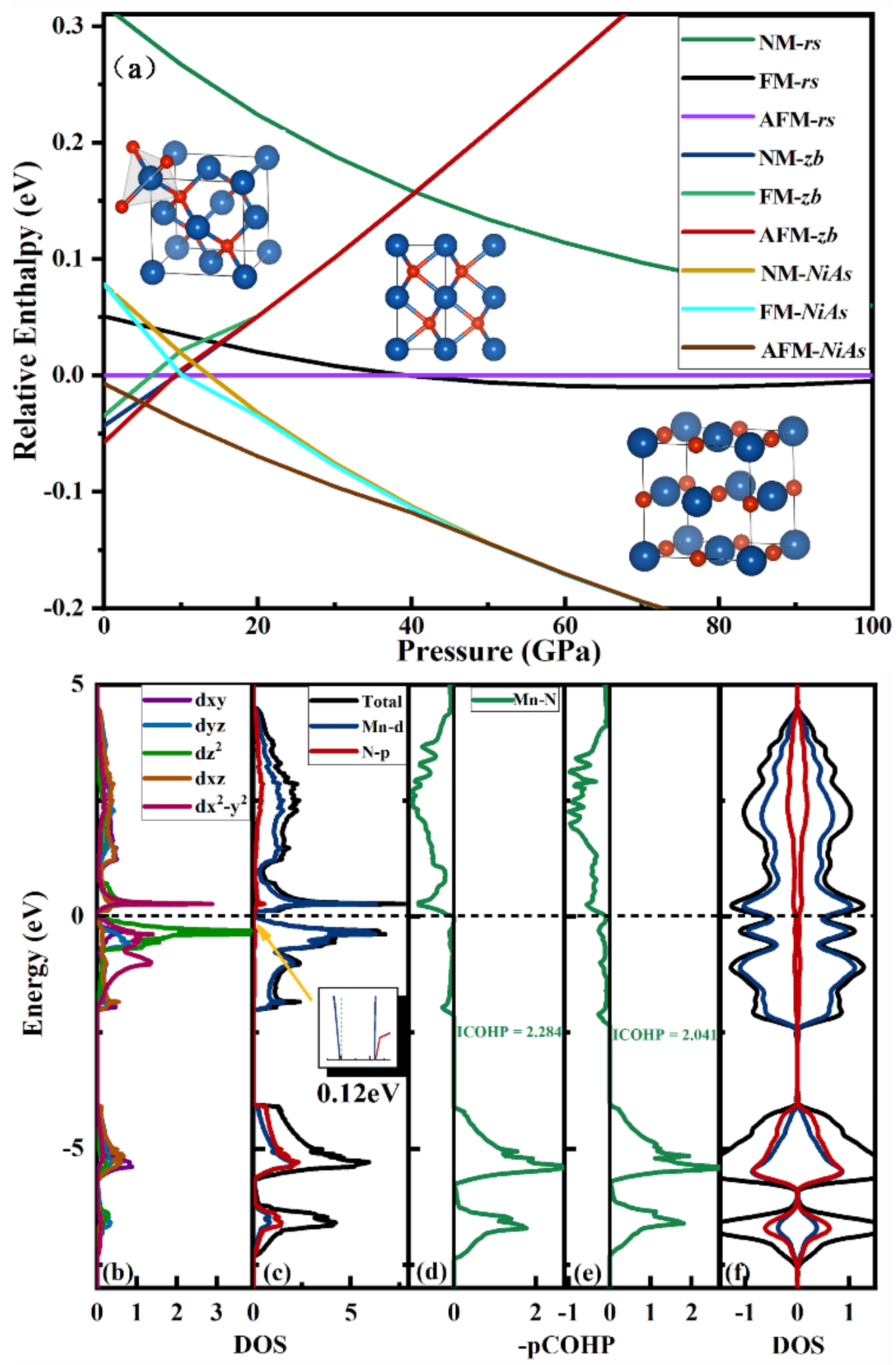

**Figure 5.** (a) The relative enthalpy of MnN and the insets are the crystal structures. (b - f) Electronic properties of *zb*-type MnN with NM and AFM phase at 0 GPa. The partial DOS and pCOHP (projected crystal orbital Hamilton population).

### 3.4. Stable Structures and Electronic Properties of $Mn_3N$, $Mn_2N$, and $Mn_3N_2$

Three Mn-rich phases have been predicted in the NM states, and they exist in hexagonal layers composed of edge-sharing $NMn_6$ tri-prisms (**Figure 6 a-c**). The difference is that the hexagonal layers are separated by the Mn layers in $Mn_3N$ and form double tri-prisms layers with increasing N content in $Mn_3N_2$. Meanwhile the crystal parameter c increased from 6.387 to 20.672 Å. The closed N-N separations are 2.587, 2.623, and 2.626 Å for the three phases and are much longer than the N-N single bond. The DOS demonstrated that these phases were remarkably metallic, and the Fermi level was dominated by the Mn-3*d* orbital. Mn-3*d* and N-2*p* exhibit the same energy peaks (−9 and −4 eV), representing a strong interaction between the N-2*p* and Mn-3*d* states. Bader charge transfer analysis reveals that each N atom of $P\bar{6}m2$-$Mn_3N$ gains 1.080 $e^-$ from the Mn atoms, and Mn loses more electrons with increasing nitrogen content in the three Mn-rich structures. This, in combination with the COHP (**Figure S7**) values, reveals that the interaction between Mn and N is enhanced with increasing nitrogen content. Moreover, there are delocalized electrons between the hexagonal layers in all the three phases (**Figure 6** d-f). This can be attributed to the metallicity of these phases.

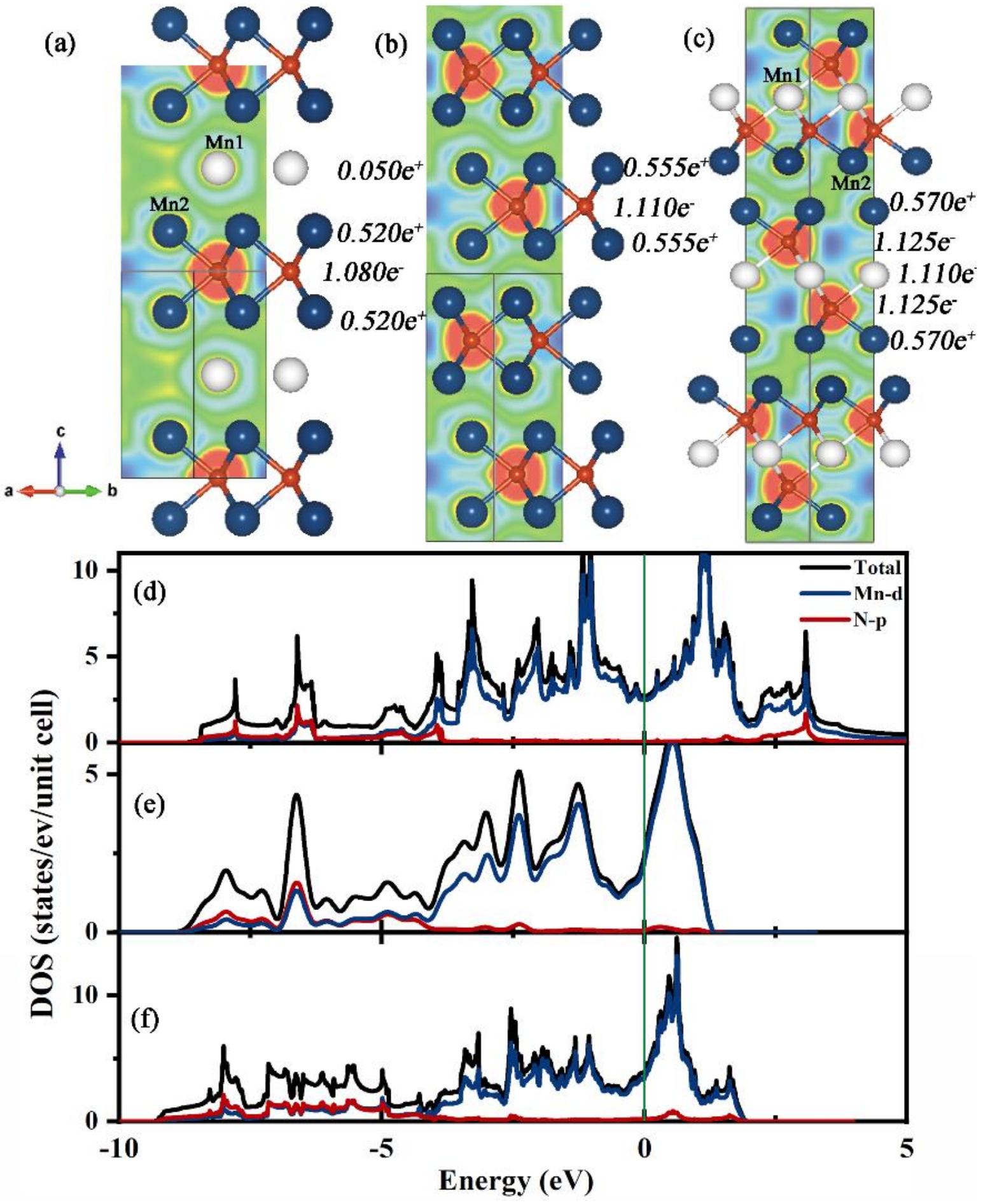

**Figure 6.** The structure diagrams, associated ELF contours (isosurface value = 0.65) and projected DOS of the (a) and (d) $P\bar{6}m2$-$Mn_3N$ at 0 GPa, (b) and (e) $P6_3/mmc$-$Mn_2N$ at 0 GPa and (c) and (f) $R\bar{3}m$-$Mn_3N_2$ at 10 GPa.

### 3.5. Mechanical Properties of Mn−N compounds

Mechanical properties determine the structural stability, and TM nitrides have always exhibited outstanding incompressibility. Thus, the elastic moduli and hardness of all the Mn−N compounds were calculated and are shown in **Figure 7a** and **Table S2**. The hardness was calculated by Chen's model (Hv1).[63] Due to minuscule G (39 GPa) in NM-*zb*-MnN, the hardness is negative. Moreover, in some TM light elements, the Chen's model may provide a hardness value that is quite different from experiment due to the lack of consideration of the metallicity and part of the covalent bond, especially in WN.[64-66] On the contrary, the results of Zhong's model (Hv2) agree well with the experiment. Thus, the two models were used in this work. *Immm*-$MnN_4$ exhibits extremely high bulk modulus B (381 GPa) and shear modulus G (213 GPa), which can be comparable to *c*-BC (B = 382 GPa)[67]. However, the hardness obtained using Chen's model is 17.5 GPa. Considering the strong covalent bond and metallicity in $MnN_4$, the hardness obtained using Zhong's model is 36 GPa, which is close to that of superhard materials (40 GPa). This can be attributed to the strong covalent bonds in the $N_4$ rings. AFM-*NiAs*-MnN is also a potential hard material with hardness above 30 GPa, based on Zhong's model. This may be attributed to the partially polar covalent bonds between Mn and N. The trend in hardness is similar to the trend in shear, G (**Figure 7**b), which changes nonlinearly with increasing nitrogen content. For the three Mn-rich phases ($Mn_3N$, $Mn_2N$, and $Mn_3N_2$), the hardness obtained by the two methods was consistent and slightly decreased with increasing nitrogen content. Thus, covalent interactions have a remarkable effect on the hardness of N−rich structures and almost no effect on Mn-rich structures in Mn−N compounds.

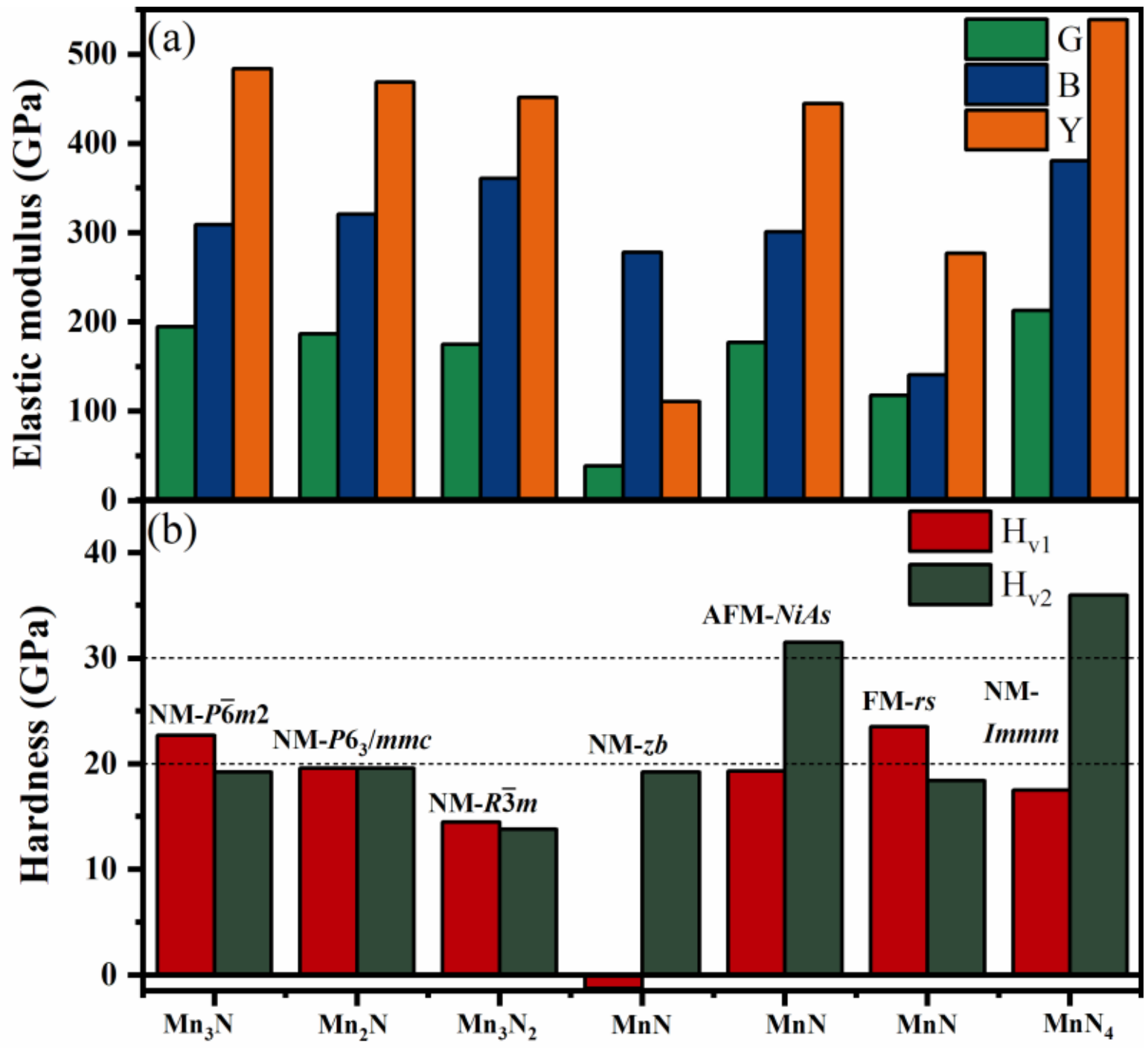

**Figure 7.** (a) Calculated bulk (B), shear (G) and Young's modulus (Y), (b) hardness for Mn−N compounds.

## Conclusions

The crystal structures of stable Mn−N compounds were investigated systematically at 0-100 GPa. A series of thermodynamically and dynamically stable Mn−N compounds were predicted for the first time. Remarkably, N-rich $MnN_4$ with a planar $N_4$ ring and $sp^2$ hybridization of N was discovered for the first time. The electron occupancy of the π-antibonding states within $N_4$ rings and the interaction between Mn and N are attributed to the metallicity of this phase, whose $T_c \approx 1.6$ K. The strong covalent bonds in the $N_4$ rings play an important role in imparting the ultra-incompressible character, which renders it a hard-superconductive material. We reconstructed the phase transition sequence for manganese mononitride. The results indicated that the semi-conducting NM-*zb* (5 GPa) phase transforms to the metallic AFM-*NiAs* (40 GPa) phase, and further transforms into the more stable metallic FM-*rs* phase. AFM-*NiAs*-MnN exhibited excellent incompressibility with a hardness higher than 30 GPa, according to Zhong's model; this enables its application in harsh environments. The mechanical properties of Mn−N compounds show that covalent interactions considerably affect the hardness of N−rich structures; however, it has almost no influence on the hardness of Mn-rich structures, which raises questions on the popular models for predicting material hardness. This study on the Mn−N system may guide the syntheses of manganese–nitrogen compounds, provide a greater understanding of their properties, and aid the application of Mn−N hard multifunctional materials.

## Acknowledgements

This work was supported by the National Key R&D Program of China (No. 2018YFA0703404, 2016YFB0201204, 2017YFA0403704), National Natural Science Foundation of China (Nos. 11774121, 91745203). Program for Changjiang Scholars and Innovative Research Team in University (No. IRT_15R23), Parts of calculations were performed in the High Performance Computing Center (HPCC) of Jilin University.

**Supporting Information**

1. Crystal structure diagrams and crystal parameters.
2. Dynamical properties of Mn-N system.
3. The relative enthalpy of $Mn_2N$ and $Mn_3N_2$.
4. The electronic properties of Mn-N system.
5. Mechanical properties of Mn-N system.

# TOC Graph

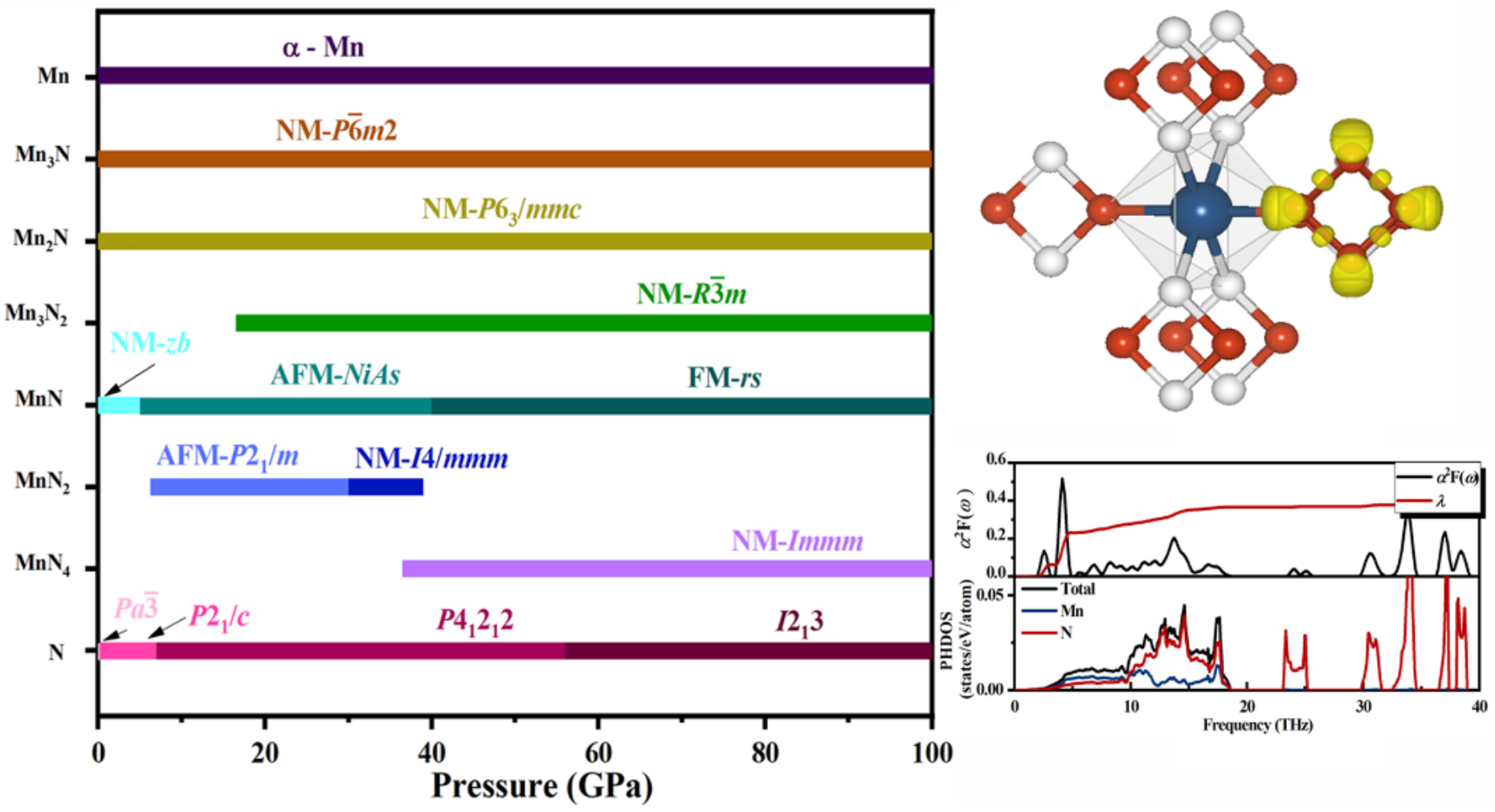